\documentclass[letterpaper]{article} 
\usepackage{aaai2026}  
\usepackage{times}  
\usepackage{helvet}  
\usepackage{courier}  
\usepackage[hyphens]{url}  
\usepackage{graphicx} 
\usepackage{natbib}  
\usepackage{caption} 
\usepackage{algorithm}
\usepackage{algorithmic}
\usepackage{amsmath}
\usepackage{amssymb}
\usepackage{booktabs}
\usepackage{array}
\usepackage{multirow}
\usepackage{multicol}
\usepackage{enumitem}
\usepackage{comment}
\usepackage{pifont} 

\usepackage{newfloat}
\usepackage{listings}
\DeclareCaptionStyle{ruled}{labelfont=normalfont,labelsep=colon,strut=off} 
\floatstyle{ruled}
\newfloat{listing}{tb}{lst}{}
\floatname{listing}{Listing}
\nocopyright
\title{Localizing Audio-Visual Deepfakes via Hierarchical Boundary Modeling}
\author{
    Xuanjun Chen\textsuperscript{\rm 1},
    Shih-Peng Cheng\textsuperscript{\rm 2}\thanks{Equal Contribution.},
    Jiawei Du\textsuperscript{\rm 2},\
    Lin Zhang\textsuperscript{\rm 3}, Xiaoxiao Miao\textsuperscript{\rm 4} \\ 
    Chung-Che Wang\textsuperscript{\rm 2}, Haibin Wu\textsuperscript{\rm 5}\thanks{Corresponding Author.}, Hung-yi Lee\textsuperscript{\rm 1}, Jyh-Shing Roger Jang\textsuperscript{\rm 2}\
}
\affiliations{
    \textsuperscript{\rm 1}Graduate Institute of Communications Engineering, National Taiwan University\\
    \textsuperscript{\rm 2}Department of Computer Science and Information Engineering, National Taiwan University\\
    \textsuperscript{\rm 3}Center for Language and Speech Processing, Johns Hopkins University \\
    \textsuperscript{\rm 4}Division of Natural and Applied Sciences, Duke KunShan University \\ \textsuperscript{\rm 5}Independent Researcher \\

}

\usepackage{bibentry}

\begin{document}

\maketitle

\begin{abstract}
Audio–visual temporal deepfake localization under the content‐driven partial manipulation remains a highly challenging task.
In this scenario, the deepfake regions are usually only spanning a few frames, with the majority of the rest remaining identical to the original.
To tackle this, we propose a Hierarchical Boundary Modeling Network (HBMNet), which includes three modules: an Audio–Visual Feature Encoder that extracts discriminative frame-level representations, a Coarse Proposal Generator that predicts candidate boundary regions, and a Fine-grained Probabilities Generator that refines these proposals using bidirectional boundary-content probabilities. 
From the modality perspective, we enhance audio–visual learning through dedicated encoding and fusion, reinforced by frame-level supervision to boost discriminability. 
From the temporal perspective, HBMNet integrates multi-scale cues and bidirectional boundary-content relationships. 
Experiments show that encoding and fusion primarily improve precision, while frame-level supervision boosts recall. 
Each module (audio-visual fusion, temporal scales, bi-directionality) contributes complementary benefits, collectively enhancing localization performance. 
HBMNet outperforms BA-TFD and UMMAFormer and shows improved potential scalability with more training data.
\end{abstract}

\section{Introduction}
\label{sec:Intro}

With the surge of generative AI, partial deepfakes are harder to catch than full deepfakes, increasing risk \cite{he2025manipulatedregionslocalizationpartially}. 
Early benchmarks on deepfake audio, such as ASVspoof~\cite{wu15e_interspeech, kinnunen17_interspeech, todisco2019asvspoof, nautsch2021asvspoof, liu2022asvspoof2021, Wang2024_ASVspoof5} and ADD~\cite{yi2022add,yi2023add} treated entire utterances as wholly bona fide or fake, mostly neglecting localized edits. 
PartialSpoof ~\cite{lin2023partialspoof} targeted word-/phrase-level forgeries but was limited in scale and modality.
Combining deepfake audio with lip-synced video yields audio-visual deepfakes that are even harder to detect. 
In the audio-visual realm, LAV-DF~\cite{ba-tfd+_cai2023glitch} pioneered content-driven localization but used limited antonym swaps and only replacement edits. AV-Deepfake-1M~\cite{cai2024av} dramatically raises the bar by combining three manipulation strategies (replacement, deletion, insertion) with frame-level annotations and high-fidelity generation and by using large language models \cite{openai2024gpt4technicalreport} to ensure diverse, context-consistent content. 
These edits demand spotting subtle multimodal inconsistencies. 
This complexity makes AV-Deepfake-1M an essential benchmark for advancing fine-grained audio-visual deepfake localization. 
The Audio-Visual Temporal Deepfake Localization (AVTDL) task seeks to localize manipulated segments in audio-visual streams. 
On AV-Deepfake-1M, uni-modal methods (e.g.,  PyAnnote \cite{plaquet23_interspeech}, Meso4 \cite{Afchar2018meso4}, EfficientViT \cite{Coccomini2022EfficientViT}, TriDet variants \cite{Shi_2023_CVPR}, ActionFormer~\cite{Zhang2022Actionformer}) fail to capture complex, asynchronous edits across modalities, whereas audio–visual approaches (e.g., BA-TFD~\cite{ba-tfd_cai2022you}, BA-TFD+~\cite{ba-tfd+_cai2023glitch}, UMMAFormer~\cite{zhang2023ummaformer}) that fuse acoustic and visual cues consistently outperform them by exploiting complementary information. 
Despite these advances, audio–visual methods still struggle with AV-Deepfake-1M’s asynchronous dependencies and fall short of accurate temporal deepfake localization. 

There are two challenges that may cause the gap. 
First, audio-visual encoding and fusion are insufficient. 
Audio and visual modalities provide complementary cues: acoustic and visual artifact features can reveal artifacts within each modality, while their interaction highlights the modality inconsistencies between real and deepfake regions. 
Despite the potential complementarity of multimodal cues, the aforementioned audio-visual methods process both streams with independent encoders, and their insufficient fusion strategies hinder true cross-modal interaction, allowing one modality to overpower the other and ultimately degrading localization performance. For example, BA-TFD and BA-TFD+ tend to rely heavily on visual cues, whereas UMMAFormer often depends on audio features. Imbalance in modality contributions ultimately limits their effectiveness, underscoring the need for more robust audio-visual integration. 

Second, there is often insufficient modeling of boundary cues in existing temporal localization. 
\textbf{1) Limited to Single Temporal Scale Boundary Cues:}
Existing audio-visual methods use either frame- or proposal-level cues and none integrates both, despite their close relationship. 
For example,  BA-TFD and BA-TFD+ rely on proposal-level features, UMMAFormer relies on frame-level features. 
Frame-level features excel at identifying precise boundaries; they pinpoint the exact start and end frames of a manipulated segment, because each frame is processed in isolation, they cannot capture broader segment-wide patterns. 
In contrast, proposal-level features aggregate information across candidate segments of varying lengths to encode global temporal context, yet this down-sampling inevitably discards critical fine-grained details. 
\textbf{2) Restricted to Unidirectional Temporal Boundary Cues:}
Prior work relies solely on a forward pass, utilizing only past context in a causal manner, making it easy to overlook post-transition information.
And, for each ``real-to-fake'' boundary transition, it misses the complementary “fake-to-real” cues that occur at the same positions.
Beyond that, they treat boundary detection and frame-wise “real versus fake” classification (content) as separate tasks, ignoring that content itself and boundary cues can mutually reinforce one another. 
None of the existing methods, including BA-TFD, BA-TFD+, or UMMAFormer, go beyond unidirectional temporal boundary cues.

To overcome these limitations, we introduce the Hierarchical Boundary Modeling Network (HBMNet), a unified framework for AVTDL task that comprises an Audio-Visual Feature Encoder (AVFE) to extract frame-level cross-modal features, a Coarse Proposal Generator (CPG) for producing a dense boundary-matching map (proposal-level), and a Fine-grained Probabilities Generator (FPG) for refining boundaries via bidirectional boundary-content probabilities (frame-level). Our key contributions are as follows:
\begin{itemize}
    \item We design HBMNet, a unified framework that jointly integrates better audio-visual learning and hierarchical boundary information, including different temporal scales (at coarse-grained proposal level and fine-grained frame level), boundary transition with temporal bi-directionality, for end-to-end localization. 

    \item From HBMNet experiments, we find that the audio–visual encoding and fusion modules work together to improve localization precision, while frame-level deepfake supervision strengthens localization recall. 
    In addition, each module (audio-visual fusion, temporal scales, bi-directionality) discussed in HBMNet delivers complementary benefits for temporal deepfake localization, collectively enhancing overall localization performance. 

    \item Extensive experiments on AV-Deepfake-1M show that HBMNet outperforms BA-TFD and UMMAFormer. HBMNet improves performance as training data increases, showing potential scalability. 

\end{itemize}

\section{Related Work}
\label{lab:related_work}

\subsection{Audio-Visual Learning}\label{sec:related_work_av}
Modeling explicit audio–visual interactions \cite{Vila2025avlearning} is crucial to capture global and local cues. We review prior work from an audio-visual learning lens:

\textbf{1) Single‐Modality Encoding:}
Early methods use independent backbones: BA-TFD/BA-TFD+ apply a 3D CNN or MViTv2 \cite{li2022mvitv2} for video and a static 2D CNN or ViT \cite{dosovitskiy2021an} for audio—overlooking long-term context \cite{liu2023audio}, while UMMAFormer relies on pretrained BYOL-A \cite{niizumi2021byol} and TSN \cite{wang2016temporal}, at significant computational cost.

\textbf{2) Cross‐Modal Fusion:}  
BA-TFD, BA-TFD+, and UMMAFormer rely on rudimentary fusion strategies. 
BA-TFD and BA-TFD+ employ late fusion by weighted sums, which limits interaction and often lets the visual stream dominate. UMMAFormer uses early fusion by concatenating audio and visual features, assuming temporal alignment and ignoring modality gaps, leading to suboptimal integration. 

\textbf{3) Frame‐Level Supervision:}
Contrastive learning \cite{Hadsell2006contrastive} is to improve feature discrimination. 
However, BA-TFD uses contrastive loss at the utterance level, lacking frame‐wise supervision. Consequently, real and manipulated frames remain insufficiently distinguished. 
Moreover, binary cross‐entropy loss ignores class imbalance, since deepfake regions occupy only a small part. 

In this study, we strengthen our audio–visual front-end by (1) efficient temporal encoders for long-range context, (2) cross-attention fusion to unite modalities, and (3) fine-grained supervision for sharper frame-level discrimination.

\begin{figure*}[t]
\centering
\includegraphics[width=17.6cm]{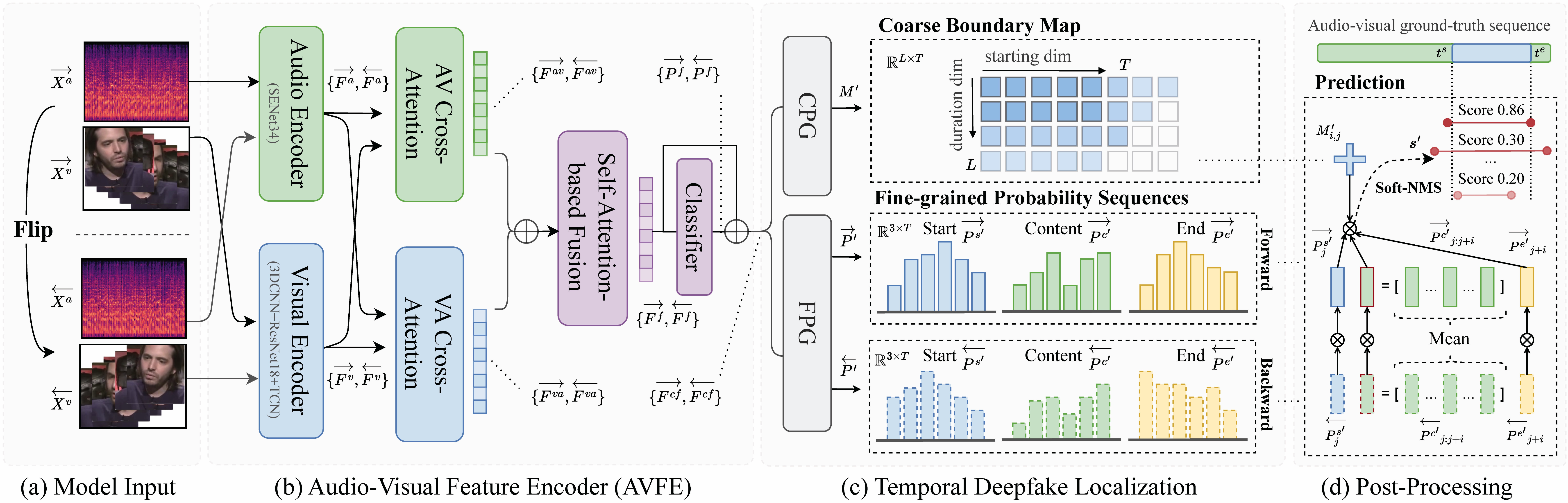}
\caption{The main stem of Hierarchical Boundary Modeling Network (HBMNet) used in training and inference. Auxiliary branch attachments share the main stem architecture but have independent weights, indicated with structures omitted for clarity.}
\label{fig:overview}
\end{figure*}

\subsection{Temporal Localization}
By revisiting the temporal localization problem in AVTDL, we categorize existing approaches along three dimensions:

\textbf{1) Audio-Visual Modalities:} 
Existing methods (BA-TFD, BA-TFD+, UMMAFormer) excel on LAV-DL \cite{ba-tfd_cai2022you, ba-tfd+_cai2023glitch} but falter on AV-Deepfake-1M, highlighting the need for stronger cross-modal interaction modeling.

\textbf{2) Different Temporal Scales:}
BA-TFD and BA-TD+ leverage proposal-level (global) cues for temporal deepfake localization, whereas UMMAFormer and its top-performing variants, Vigo \cite{1M-deepfakes-vigo-first} and Maya \cite{1M-deepfakes-mfms-second} from the AV-Deepfake-1M Challenge 2024, rely on frame-level (local) cues. To date, no previous work has better used global-local boundary cues for AVTDL. 

\textbf{3) Bidirectional Boundary-Content Relationships:}
Methods from the temporal action localization field \cite{Wang2024tal}, notably BSN++ \cite{BSN++_Su_Gan_Wu_Qiao_Yan_2021}, which applies bidirectional temporal boundary matching, and MCBD \cite{su2023multi}, which adds content-aware supervision. 
Bidirectional boundary–content relationships improve performance in visual-only settings, but their effectiveness in audio–visual localization remains unverified. 

In this work, we propose HBMNet to (1) leverage hierarchical boundary cues within a unified framework (global–local, bidirectional boundary-content, audio–visual) and (2) show their complementarity via systematic studies.

\section{Proposed Approach}

\subsection{Problem Formulation}
We formally define the AVTDL problem as follows. Let  
$X^a$ and $X^v$   
represent the audio and visual inputs. 
The ground truth is denoted as  
$\Psi = \{\varphi_n = (t^s_n, t^e_n)\}_{n=1}^{N}$,  
where $N$ is the number of segments. 
Here, $\varphi_n$ represents the label for the $n$-th segment, with $t^s_n$ and $t^e_n$ indicating the time of its start and end boundaries. 
These segments may contain partial manipulations in either modality. A segment is labeled as ``real'' if and only if both its audio and visual modalities remain unmodified within $\{t^s_n, t^e_n\}$; otherwise, it is labeled as ``fake.''  
$\Psi' = \{\varphi_n' = (t'^s_n, t'^e_n)\}_{n=1}^{N'}$  
is used to represent a set of proposals, where $t'^s_n$ and $t'^e_n$ denote the predicted start and end times. Performance is evaluated based on how well $\Psi'$ captures the ground truth regions $\Psi$, prioritizing high recall and strong temporal overlap between the predicted and ground-truth regions.
Besides, ground-truth for boundary-matching confidence map ($M$) in the proposal-level and for start/end/content in the frame-level ($P$) and are also derived from $\Psi$ for training which will be introduced in Section~\ref{sec:training}.

\subsection{Motivation}
As introduced in Section~\ref{sec:Intro}, audio-visual deepfake localization faces two main challenges: (1) insufficient audio-visual learning and (2) lack of boundary cues. The proposed HBMNet aims to mitigate these challenges from two points. 

\textbf{1) Robust audio-visual encoding and fusion.}
As claimed in \cite{cai2024av}, 
deepfakes across different modalities can cause misalignment between them, which could aid in locating deepfake regions. Thus, modeling cross-modal audio-visual information is crucial. 
To this end, our Audio–Visual Feature Encoder incorporates: dedicated temporal encoders to capture long-range dependencies, a cross‐attention fusion module to bridge audio–visual gaps, and fine‐grained supervision frame‐level contrastive loss to sharpen real‐vs‐fake discrimination. 
This combination ensures the encoder produces rich representations. 

\textbf{2) Hierarchical boundary modeling.}
HBMNet unifies multi-scale and bidirectional boundary cues in an end-to-end framework. 
It leverages temporal scale, frame-level versus proposal-level, to decide when to focus on fine or coarse context, and bidirectional encoding, forward and backward passes, to decide how to interpret transition signals. First, a coarse boundary map over the entire sequence captures global context; then, dual-pass encodings produce three frame-wise probability streams (start, end, and content), integrating multi-scale detail with complementary forward and backward cues. 
This combination of different temporal scales information and bidirectional transitions enables precise localization of even the smallest manipulated fragments.

\subsection{Hierarchical Boundary Modeling Network} \label{sec:HBMNet}

The main stem of HBMNet is shown in Figure~\ref{fig:overview}. In phases (a) and (b), an Audio-Visual Feature Encoder (AVFE) encodes and fuses audio and visual sequences to capture cross-modal temporal dependencies. In phase (c), the fused features split into two heads: the Coarse Proposal Generator (CPG), which produces a dense boundary map, and the Fine-grained Probabilities Generator (FPG), which outputs start, end, and content probabilities. Finally, phase (d) integrates these outputs to localize manipulated segments precisely.

\subsubsection{Bi-directional Input.} In phase (a), to capture bidirectional boundary-content cues, HBMNet processes both original and reversed inputs. We denote the original inputs as $\overrightarrow{\mathbf{X}}^a$ and $\overrightarrow{\mathbf{X}}^v$ and their flipped versions as $\overleftarrow{\mathbf{X}}^a$ and $\overleftarrow{\mathbf{X}}^v$. By utilizing bidirectional input, the model avoids missing crucial post-transition information and learns additional fake-to-real transitions for every real-to-fake transition, and vice versa, enhancing frame-level discriminability.

\subsubsection{Audio-Visual Feature Encoder.} \label{subsubsec: AVFE}

AVFE aims to extract $\{\overrightarrow{F^{cf}}, \overleftarrow{F^{cf}}\}$ from the bidirectional input sequences of audio and visual modalities into a shared feature space. As shown in Figure~\ref{fig:overview}-(b), AVFE consists of two stages: single-modality encoding and cross-modality encoding.
In the first single-modality encoding stage, the audio encoder and visual encoder separately convert the input sequence $X$ into per-frame feature sequences in their respective spaces, $\{\overrightarrow{F^a}, \overleftarrow{F^a}\}$ for audio and $\{\overrightarrow{F^v}, \overleftarrow{F^v}\}$ for visual input, each processed in both directions.
Specifically, input audio is first converted to mel-spectrograms and then fed into SENet~\cite{hu2018senet} as the audio encoder. And, the visual encoder employs a 3D CNN stem followed by a ResNet-18 ~\cite{he2016resnet} and TCN modules~\cite{martinez2020lipreading}. 

In the second stage, cross-attention is applied to capture temporal dependencies across modalities from audio $\{\overrightarrow{F^a}, \overleftarrow{F^a}\}$ and visual features $\{\overrightarrow{F^v}, \overleftarrow{F^v}\}$. Two types of attention mechanisms are applied: cross-attention and self-attention. The cross-attention structure has been shown to be effective in audio-visual learning tasks~\cite{chen2023push, chen2024multi}. Following their approach, we utilize three attention blocks to merge audio and visual features: audio-to-visual (AV), visual-to-audio (VA), and a self-attention-based fusion (AttFUS) block.
Specifically, the AV and VA blocks leverage cross-attention mechanisms~\cite{vaswani2017attention} to exchange cross-modal information by treating audio and visual features as queries alternately. 
This process extracts $\{\overrightarrow{F^{av}}, \overleftarrow{F^{av}}\}$ and $\{\overrightarrow{F^{va}}, \overleftarrow{F^{va}}\}$. The fusion block then aggregates these outputs via self-attention to obtain the fused feature set $\{\overrightarrow{F^f}, \overleftarrow{F^f}\}$.
Consequently, $\{\overrightarrow{F^f}, \overleftarrow{F^f}\}$ 
are passed into a classifier to produce a frame-wise probability sequence $P^f \in \mathbb{R}^T$. 
Finally, the outputs $\{\overrightarrow{F^f}, \overleftarrow{F^f}\}$ and $\{\overrightarrow{P^f}, \overleftarrow{P^f}\}$ are concatenated as $\overrightarrow{F^{cf}} = \overrightarrow{F^f} \oplus \overrightarrow{P^f}$ and $\overleftarrow{F^{cf}} = \overleftarrow{F^f} \oplus \overleftarrow{P^f}$, and passed to the subsequent CPG and FPG modules. 

\subsubsection{Coarse Proposal Generator.}\label{sec:CPG}

CPG employs a Boundary-Matching (BM) layer~\cite{BMN_Lin_2019_ICCV} to convert the forward-direction feature $\overrightarrow{F^{cf}} \in \mathbb{R}^{C \times T}$ into a confidence map $M' \in \mathbb{R}^{L \times T}$ (see Figure~\ref{fig:overview}-(c), top), where $C$ is the feature dimension, $T$ the number of start times, and $L$ the maximum proposal duration. Each entry $M'_{i,j}$ scores the proposal $\varphi_{i,j} = (t^s = t_j,\ t^e = t_j + t_i)$. 
To compute $M'$, the BM layer constructs a sampling mask $W \in \mathbb{R}^{N \times T \times L \times T}$, where $N$ is the number of uniformly sampled points in $[t^s, t^e]$. Each mask $w_{i,j} \in \mathbb{R}^{N \times T}$ softly samples $\overrightarrow{F^{cf}}$ to yield a feature of shape $C \times N$. Aggregating all $\varphi_{i,j}$ results in a BM feature map $M_F \in \mathbb{R}^{C \times N \times L \times T}$.
This map is further processed by 3D and 2D convolutions to produce $M'$, capturing contextual cues across proposals of varying durations. This unified modeling enables CPG to assess temporal boundaries from a global perspective.

\subsubsection{Fine-grained Probabilities Generator.}
The goal of FPG is to  
produce fine-grained \textit{bidirectional} boundary-content probabilities. 
While the Coarse Proposal Generator (CPG) provides a global boundary overview, its reliance on sampling to construct boundary maps over proposals of varying lengths inevitably sacrifices information. 
FPG complements CPG by generating detailed frame‐level start, end and content probabilities in both forward and backward, thereby providing the fine‐grained cues missing in CPG module. 

FPG fully leverages bidirectional cues by fusing forward and backward predictions via element-wise geometric mean.
As illustrated in Figure~\ref{fig:overview}-(c), FPG models the likelihood of each temporal position being a start, end, or content region, through bidirectional probability sequences. This bidirectional design helps disambiguate boundaries and capture asymmetric transitions, thereby improving performance.

FPG based on Nested U-Net~\cite{zhou2018unet++}, taking the bidirectional fusion features $\{\overrightarrow{F}^{cf}, \overleftarrow{F}^{cf}\}$ as input.
FPG downsamples them through two pooling stages with interleaved convolutional units, then upsamples via skip connections. 
The output of FPG yield the bidirectional fine-grained probability predictions $\overrightarrow{P'} = \{\overrightarrow{{P^s}'}, \overrightarrow{{P^e}'}, \overrightarrow{{P^c}'}\}$ and $\overleftarrow{P'} = \{\overleftarrow{{P^s}'}, \overleftarrow{{P^e}'}, \overleftarrow{{P^c}'}\}$, where $\overrightarrow{P'}$ and $\overleftarrow{P'} \in \mathbb{R}^{3 \times T}$ correspond to three dimensions derived from different channels of the convolutional layers. These dimensions represent the predicted probability sequences for (1) whether certain temporal locations are the start or end boundaries, and (2) whether specific contents are fake. 

\subsection{Model Training}\label{sec:training}
The overall training objective comprises three terms: a frame-level contrastive loss $\mathcal{L}_{\mathrm{FC}}$ on AVFE and two temporal deepfake localization losses $\{\mathcal{L}_{\mathrm{CP}}, \mathcal{L}_{\mathrm{FP}}\}$ applied to CPG and FPG
The overall loss function $\mathcal{L}_{\mathrm{HBMNet}}$, is defined as:
\begin{equation}
\begin{aligned}
\mathcal{L}_{\mathrm{HBMNet}} = 
\alpha \cdot \mathcal{L}_{\mathrm{FC}}
+ \mathcal{L}_{\mathrm{CP}}
+ \mathcal{L}_{\mathrm{FP}},
\end{aligned}
\end{equation}
where $\alpha$ is the weight of the frame-level contrastive loss. 

\subsubsection{Frame-Level Contrastive Loss on AVFE.}\label{sec:avfe_loss}
To enforce audio–visual consistency, we apply a frame-level contrastive loss $\mathcal{L}_{\mathrm{FC}}$ to the hidden embeddings $F^{av} = \{\overrightarrow{F^{av}}, \overleftarrow{F^{av}}\}$, 
$F^{va} = \{\overrightarrow{F^{va}}, \overleftarrow{F^{va}}\}$ on AVFE.

\textit{Label Construction:}  
We construct a label sequence $Y = \{y_{t}\}_{t=1}^T \in \mathbb{R}^T$ at the uniform frame level, where $T$ is the total number of frames and $y_{t}$ is the label for the $t$-th frame. A frame $y_{t}$ is labeled as ``real'' if and only if both audio and visual of the $t$-th frame are real; Otherwise, it’s labeled ``fake.''  

\textit{Loss Calculation:} 
We adopt frame-level contrastive learning $\mathcal{L}_{\mathrm{FC}}$ ~\cite{Hadsell2006contrastive}, which encourages the discriminability of audio-visual features.  
The loss $\mathcal{L}_{\mathrm{FC}}$ is designed to pull closer the cross-modal features of real frames while pushing apart those of manipulated frames, and is defined as follows:
\begin{equation}
\mathcal{L}_{\mathrm{FC}} = \frac{1}{T} \sum_{t=1}^{T} \mathrm{Contrastive}(d_t, y_t),
\end{equation}
where $\mathrm{Contrastive}(\cdot)$ is a margin-based contrastive loss. 
The cross-modal \( L_2 \) distance $d_t$ at frame $t$ is defined as:
\begin{equation}
d_t = \| \overrightarrow{F^{av}_t} - \overrightarrow{F^{va}_t} \|_2 + \| \overleftarrow{F^{av}_t} - \overleftarrow{F^{va}_t} \|_2.
\end{equation}

\subsubsection{Deepfake Localization Loss on CPG}
The loss $\mathcal{L}_{\mathrm{CP}}$ guides the predicted proposals to match the ground-truth deepfake proposals at the proposal level. 
\textit{Label Construction:}  
We construct a label matrix $M \in [0,1]^{L\times T}$. 
where each element $M_{i,j}$ denotes the proportion of the region $(t_j,t_j+t_i)$ that overlaps with the most relevant fake region and is labeled as fake.
Here, $L$ denotes the maximum number of all proposal durations, and $T$ represents the total number of possible temporal start times.  
Each $M_{i,j}$ is calculated based on its anchor proposal $\varphi^a_{i,j}$, with $t_j$ as the start time and $t_j + t_i$ as the end time. Specifically, for each anchor proposal $\varphi^a_{i,j}$, we compute its Intersection-over-Union (IoU) with all ground-truth regions $\varphi \in \Psi$ and assign to $M_{i,j}$ the maximum IoU obtained:
\begin{equation}
M_{i,j} = \max_{\varphi \in \Psi} \mathrm{IoU}\bigl(\varphi^a_{i,j}, \varphi \bigr).
\end{equation}
\textit{Loss Calculation:} Given the fused feature $F^{cf}$, CPG predicts a coarse boundary map $M'$ as introduced in Section~\ref{sec:CPG}. 
The loss $\mathcal{L}_{\mathrm{CP}}$ is then calculated using the mean squared error (MSE) between the predicted boundary map $M'$ and the ground truth boundary map $M$:
\begin{equation}
\mathcal{L}_{\mathrm{CP}} = \mathrm{MSE}(M', M).
\label{equation:cpg}
\end{equation}
This encourages CPG to provide a rough localization by jointly modeling temporal boundaries across diverse proposals, while leveraging the dense proposal space to capture structural relationships for localization.

\subsubsection{Deepfake Localization Loss on FPG} The $\mathcal{L}_{FP}$ loss on FPG aims to model the frame-level boundary detection.

\textit{Label Construction:}  
Similar to BSN~\cite{BSN_Lin_2018_ECCV}, we define the ground-truth temporal region for a time point $t$ as $r(t) = \bigl[t - \tfrac{d^f}{2}, t + \tfrac{d^f}{2}\bigr]$, where $d^f$ is a
predefined interval with fixed duration.  
Given a ground-truth proposal $\varphi = (t^s, t^e) \in \Psi$, the ground-truth temporal regions of starting and ending points are defined as: $r^s = r(t^s), \quad r^e = r(t^e)$. The ground-truth region of content is defined as: $r^c = \varphi$. 

To obtain uniform temporal anchor labels (i.e., start, end, and content), we define anchors $a_t$ and compute the Intersection over Anchor (IoA) between each anchor and the ground-truth regions. 
For each anchor $a_t = [a_t^{\min}=d^f \cdot (t-0.5), a_t^{\max}=d^f \cdot (t+0.5)]$ at time step $t$, we take the maximum IoA between the anchor and the ground-truth regions:
$p_t^s = \max_{k} \text{IoA}(a_t, r_k^s), \quad p_t^e = \max_{k} \text{IoA}(a_t, r_k^e), \quad p_t^c = \max_{k} \text{IoA}(a_t, r_k^c)$,
where $\text{IoA}(a_t, r_k) = \frac{\text{dur}(a_t \cap r_k)}{\text{dur}(a_t)}$ measures the overlap ratio between anchor $a_t$ and ground-truth region $r_k$. 
Finally, we get start, end, and content label sequences:
$
\bigl\{P^b\bigr\}_{b\in\{s,e,c\}}
\;=\;\bigl\{\{\,p^b_t\}_{t=1}^T \;\bigm|\;b\in\{s,e,c\}\bigr\}.
$

\textit{Loss Calculation:} 
Given bidirectional fine-grained probability predictions for start, end, and content 
$\overrightarrow{P'} = \{\overrightarrow{{P^s}'}, \overrightarrow{{P^e}'}, \overrightarrow{{P^c}'}\}$ and $\overleftarrow{P'} = \{\overleftarrow{{P^s}'}, \overleftarrow{{P^e}'}, \overleftarrow{{P^c}'}\}$
produced from FPG, we utilize Focal loss \cite{lin2017focal} to define the training objective for better enhancing the model's focus on hard-to-predict frames. The Focal loss is defined as:
\begin{equation}
\text{Focal}(P', P) = -\beta (1 - P)^{\beta_1} \text{BCE}(P', P)
\end{equation}
where $\sigma(\cdot)$ is the sigmoid activation function, $\beta = \beta_0 \cdot P + (1-\beta_0) \cdot (1-P)$ is the balanced weight for class imbalance, $\beta$ is the focusing parameter that down-weights easy examples, $\text{BCE}(\cdot)$ 
The complete training objective is defined as:
\begin{equation}
\mathcal{L}_{\mathrm{FP}} = \sum_{b \in \{s, e, c\}} \text{Focal}(\overrightarrow{P}^{b'}, \overrightarrow{P}^{b}) 
+ \text{Focal}(\overleftarrow{P}^{b'}, \overleftarrow{P}^{b}),
\label{equation:fpg}
\end{equation}
where $b \in \{s, e, c\}$ denotes start, end, and content probabilities, respectively. 
The loss trains HBMNet on learns bidirectional temporal dependencies for boundary localization.

\subsection{Inference and Post-Processing}
During inference, CPG outputs a coarse boundary-matching map $M' \in \mathbb{R}^{L \times T}$, where each element $M'_{i,j}$ represents the confidence of a candidate proposal $\varphi_{i,j}$ with starting frame $j$ and duration $i$. 
Additionally, FPG produces bidirectional fine-grained probability sequences for start, end, and content. 
Let $P'=\{P^{s'}, P^{e'}, P^{c'}\} \in \mathbb{R}^{3\times T}$ denote the start, end, and content probabilities. 
Both forward and backward predictions are computed, denoted as $\overrightarrow{P'}$ and $\overleftarrow{P'}$. 
The bidirectional predictions are fused via geometric-mean fusion:
\begin{equation}
    \overleftrightarrow{P'}=\sqrt{\overrightarrow{P'} \times \overleftarrow{P'}},
\end{equation}
where the addition is element-wise within each of $P^s, P^e, P^c$. 
For each candidate proposal $\varphi_{i,j}$, the final confidence score $\varphi_{i,j}$ is computed as:
\begin{equation}
    s' = M'_{i,j} \times \overleftrightarrow{P}^{s'}_{j} \times \overleftrightarrow{P}^{e'}_{j+i} \times \text{mean}(\overleftrightarrow{P}^{c'}_{j:j+i}).
\end{equation}
The fused score captures both global proposal confidence and local boundary and content cues. Based on these scores, we construct a set of scored proposals $\Psi' = \{\varphi'_n = (t^s_n,\,t^e_n, s'_n)\}_{n=1}^{N'}$, where each proposal is associated with a confidence score indicating its likelihood of being a manipulated segment. 
To obtain a compact set of predictions, Soft-NMS~\cite{Bodla2017soft-nms} is applied to suppress redundancies by decaying the scores of overlapping candidates, and the top-ranked proposals are retained as the final prediction set.

\section{Experimental Setup}
The experiments are performed on AV-Deepfake-1M benchmark\footnote{https://github.com/ControlNet/AV-Deepfake1M}, which comprises 2,068 subjects, 286,721 real videos, and 860,039 deepfake videos. Its content manipulation scenarios include (1) video-only, (2) audio-only, and (3) combined audio-video. Note that a video can contain multiple deepfake regions, which is more challenging. Due to limited GPU resources, we sampled a subset from AV-Deepfake1M for experiments; the subset has 8,000 videos for training, 1,000 for validation, and 2,000 for testing. 
We sample data using a fixed random seed to ensure reproducibility.
See Appendix for dataset and HBMNet model configuration details.

For HBMNet main configuration, the maximum segment length $L$ of in CPG is 60, and its maximum frames $T=512$. 
The hyperparameter $\alpha$ of the loss function is set to 0.1. 
The Adam optimizer starts with a learning rate of $10^{-4}$, halving it if the validation loss does not decrease for three epochs. Training lasts 40 epochs with a batch size of 8. All experiments are conducted using an NVIDIA Tesla V100 GPU. 
We reproduce BA-TFD\footnote{https://github.com/ControlNet/LAV-DF} and UMMAFormer\footnote{https://github.com/ymhzyj/UMMAFormer} baseline models based on their official repositories for comparison. 

Following previous works \cite{cai2024av}, we use Average Precision (AP) and Average Recall (AR) as evaluation metrics for temporal deepfake localization. For AP, Intersection over Union (IoU) thresholds are set at 0.5, 0.75, and 0.95.
A higher threshold indicates a stricter criterion for overlap between the predicted and the deepfake segments, implying a stringent distinction in fake localization. 
AR is computed by evaluating a fixed number of proposals (50, 20, and 10) across IoU thresholds ranging from 0.5 to 0.95 with a step size of 0.05, considering the number of deepfake segments. 
For clarity, we denote AP@0.5, AP@0.75, and AP@0.95 as the AP scores under different IoU thresholds, and AR@50, AR@20, and AR@10 as the AR scores under different numbers of proposals throughout the discussion. 

\begin{table}[t]
    \centering
    \fontsize{9}{11}\selectfont
    \setlength{\tabcolsep}{1mm}
    \begin{tabular}{>{\centering\arraybackslash}m{2.5cm}|
    >{\centering\arraybackslash}m{1.7cm}>{\centering\arraybackslash}m{1.7cm}>{\centering\arraybackslash}m{1.7cm}}
        \toprule
        \multirow{2}{*}{Method}  & \multicolumn{3}{c}{Avg. Precision \% ($\pm$ Variance)} \\
                                  & 0.5 & 0.75 & 0.95 \\
        \midrule
        BA-TFD  & 38.89 $\pm$ 2.18 & 11.84 $\pm$ 0.74 & 0.10 $\pm$ 0.02  \\
        UMMAFormer & 75.93 $\pm$ 0.16 & 60.88 $\pm$ 1.17 & 9.21 $\pm$ 1.57  \\
        HBMNet (Our) & \textbf{96.82} $\pm$ 0.99 &  \textbf{90.84} $\pm$ 1.34  & \textbf{12.79} $\pm$ 0.46 \\
        
        \toprule
        \multirow{2}{*}{Method} & \multicolumn{3}{c}{Avg. Recall \% ($\pm$ Variance)} \\
                      &  50 & 20 & 10 \\
        \midrule
        BA-TFD  & 42.56 $\pm$ 2.70 & 36.30 $\pm$ 2.70 & 32.02 $\pm$ 2.83 \\
        UMMAFormer  & 78.31 $\pm$ 0.17 & 74.19 $\pm$ 0.42 & 70.68 $\pm$ 0.36 \\
        HBMNet (Our)  & \textbf{88.83} $\pm$ 0.27 & \textbf{88.49} $\pm$ 0.21 & \textbf{88.29} $\pm$ 0.26 \\
        \bottomrule
    \end{tabular}
    \caption{Main results compared to prior models}
    \label{table:SoTAmodel_comparison}
\end{table}

\section{Results and Discussions}

\subsection{Does HBMNet work effectively?}
\label{sec:hbmnet_work_effectively}
We compare HBMNet with baselines BA-TFD and UMMAFormer in Table~\ref{table:SoTAmodel_comparison}\footnote{To ensure the reliability, we resampled the dataset 3 times for training and testing, and reported the variance of the evaluation.}.
HBMNet achieves 96.82 AP@0.5, outperforming UMMAFormer (75.93) and BA-TFD (38.89). For AR@50, HBMNet also leads with a score of 88.83, compared to 78.31 for UMMAFormer and 42.56 for BA-TFD. BA-TFD’s performance collapses at higher thresholds, with AP@0.95 approaching zero. Overall, HBMNet consistently surpasses both baselines across precision and recall metrics. 
Unlike BA-TFD (proposal-only) and UMMAFormer (frame-only), HBMNet learns richer audio-visual representations and leverages hierarchical boundary cues, from coarse proposals to fine-grained probabilities, to achieve more precise boundary localization.

\subsection{
How each module aids audio-visual learning? \\
} \label{subsec:av_learning}

We compare different strategies with BA-TFD baseline to investigate the impact on audio-visual learning in Table~\ref{table:ablation_study_feature_learning}. 
We focus on audio-visual learning and keep the localization module unchanged (only CPG is used, without using FPG), following BA-TFD. 
Row 1 is the result of baseline BA-TFD model.
Then, we swap BA-TFD’s 2D/3D CNN encoders for our HBMNet AVFE Encoder (Row 2), boosting AP@0.5 by 55.65\% and AP@0.75 by 46.2\%, with AR@50/20/10 also up across the board. 
Next, adding frame-level supervision via contrastive and focal losses (Row 3) yields the highest recall. 
Finally, replacing naïve feature concatenation with our attention-based fusion (Row 4 vs. Row 3) raises AP\@0.75 from 62.3 to 69.8, confirming the benefit of deeper cross-modal interaction. 
In summary, improvements to single-modality encoders and cross-attention-based feature fusion contribute to higher localization precision, whereas frame-level contrastive loss significantly enhances the recall. 
The detailed BA-TFD vs. HBMNet comparison is in Appendix.

\begin{table}[t]
    \centering
    \fontsize{9}{11}\selectfont
    \begin{tabular}{
    >{\centering\arraybackslash}m{0.29cm}
    >{\centering\arraybackslash}m{0.29cm}
    >{\centering\arraybackslash}m{0.29cm}>{\centering\arraybackslash}m{0.29cm}
    >{\centering\arraybackslash}m{0.51cm}
    >{\centering\arraybackslash}m{0.51cm}>{\centering\arraybackslash}m{0.51cm}>{\centering\arraybackslash}m{0.51cm}
    >{\centering\arraybackslash}m{0.51cm}>{\centering\arraybackslash}m{0.51cm}>{\centering\arraybackslash}m{0.51cm}}
        \toprule
         \multirow{2}{*}{Enc.} & \multirow{2}{*}{Sup.} & \multicolumn{2}{c}{Fusion} & \multicolumn{3}{c}{Avg. Precision \%} & \multicolumn{3}{c}{Avg. Recall \%} \\
         \cmidrule(lr){5-7} \cmidrule(lr){8-10}
         
          &  & Con & Our & 0.50  & 0.75 &  0.95 &  50  &  20  &  10  \\
         
        \midrule
        O & O & O & { N} & 38.89 & 11.84 & 0.10 & 42.56 & 36.63 & 32.02  \\
         
         { Y} & O & O  & { N} & 94.54 &  58.05  & 0.11  & 69.03  & 66.56  & 65.80  \\
        
         { Y} & { Y}  & O & { N} &  94.68 & 62.30 & 0.36 & \textbf{70.95} & \textbf{68.91} & \textbf{68.10} \\

         { Y} &  { Y} & { N} & { Y} & \textbf{95.81} & \textbf{69.82} & \textbf{0.66} & 70.40 & 68.44 & 67.61 \\ 

        \bottomrule
    \end{tabular}
    \caption{Ablation study of the base module. O: original BA-TFD setting; Y: proposed HBMNet component; N: not included. Bold mark top-1 performances.}
    \label{table:ablation_study_feature_learning}
\end{table}

\begin{table}[t]
\centering
\fontsize{9}{11}\selectfont
\setlength{\tabcolsep}{1mm}
\begin{tabular}{
>{\centering\arraybackslash}m{1.2cm}>{\centering\arraybackslash}m{1.2cm}>{\centering\arraybackslash}m{0.7cm}>{\centering\arraybackslash}m{0.7cm}>{\centering\arraybackslash}m{0.7cm}>{\centering\arraybackslash}m{0.7cm}>{\centering\arraybackslash}m{0.7cm}>{\centering\arraybackslash}m{0.7cm}}
    \toprule
     \multirow{2}{*}{Visual}  & \multirow{2}{*}{Audio}  &  \multicolumn{3}{c}{Avg. Precision \%} &
     \multicolumn{3}{c}{Avg. Recall \%} \\
    \cmidrule(lr){3-5} \cmidrule(lr){6-8}
           &        &  0.5  & 0.75  & 0.95 &  50    & 20     & 10 \\
    \midrule
    Y & N & 65.56 & 57.64 & 5.30 & 61.35  &  60.21 &  59.53  \\
    N & Y &  63.57  & 61.47 &  8.60  &  63.01  &  61.14  &  60.47 \\
    Y & Y &  \textbf{96.50}  &  \textbf{92.74}  &  \textbf{12.71}  &  \textbf{89.19} &  \textbf{88.71}  & \textbf{88.29} \\
\bottomrule
\end{tabular}
\caption{Ablation study of audio-visual modalities. }
\label{table:model_trained_on_different_modalities}
\end{table}

\subsection{
How each module aids localization?
}\label{sec:each_module_loc}
HBMNet leverages audio-visual modalities, different temporal scales, and bidirectional boundary-content cues. This subsection examines how each module aids localization.

\noindent\textbf{1) Audio-visual modalities.}
We conduct an audio and visual modalities ablation study using the proposal HBMNet in Table~\ref{table:model_trained_on_different_modalities}. The results clearly demonstrate that multimodal yields the best results. Although models using only audio perform slightly better than those using only visual input, both underperform, indicating that either modality alone is insufficient for better performance. These findings highlight the complementary nature of audio and visual modalities and demonstrate that HBMNet effectively leverages this complementarity for audio-visual deepfake localization.

\noindent\textbf{2) Different temporal scales.}
We compare combinations of \texttt{CPG} and \texttt{FPG} during training and inference in Table~\ref{table:model_trained_on_different_supervision}.
Models trained with only \texttt{CPG} achieve high precision at lower IoU thresholds (e.g., 95.81 AP@0.5) but struggle at stricter thresholds, indicating limited boundary precision. 
Models trained with only \texttt{FPG} perform better at higher IoU thresholds (e.g., 9.17 AP@0.95) and achieve the highest recall, suggesting that frame-level cues enhance localization recall. 
Combining \texttt{CPG} and \texttt{FPG} during both training and inference yields the best results, with the ``\texttt{Both-Both}'' configuration achieving the highest precision and consistently strong recall. We confirm that both proposal- and frame-level information are critical for complementary integration.
\begin{table}[t]
\centering
\fontsize{9}{11}\selectfont
\setlength{\tabcolsep}{1mm}
\begin{tabular}{
>{\centering\arraybackslash}m{0.8cm}>{\centering\arraybackslash}m{1.1cm}>{\centering\arraybackslash}m{0.8cm}>{\centering\arraybackslash}m{0.8cm}>{\centering\arraybackslash}m{0.8cm}>{\centering\arraybackslash}m{0.8cm}>{\centering\arraybackslash}m{0.8cm}>{\centering\arraybackslash}m{0.8cm}}
    \toprule
     \multirow{2}{*}{Train}  & \multirow{2}{*}{Inference}  &  \multicolumn{3}{c}{Avg. Precision \%} &
     \multicolumn{3}{c}{Avg. Recall \%} \\
    \cmidrule(lr){3-5} \cmidrule(lr){6-8}
        &      &  0.5 & 0.75 & 0.95 &  50 & 20 & 10 \\
    
    \midrule
    CPG  & CPG & 95.81 & 69.82  & 0.66  &  70.40  &  68.44  & 67.61 \\
    FPG  & FPG & 78.60 & 73.32  & 9.17  &  87.93  &  87.36  & 86.58  \\
    Both & CPG & 95.95 & 81.13  & 1.88  & 76.45  &  75.15  & 74.72  \\
    Both & FPG & 81.33 & 75.05  &  9.36 &  88.06   &  87.54  &  86.72  \\
    
    Both  & Both &  \textbf{96.50}  &  \textbf{92.74}  &  \textbf{12.71} & \textbf{89.19} &  \textbf{88.71} & \textbf{88.29} \\
\bottomrule
\end{tabular}
\caption{Ablation study of CPG and FPG modules.}
\label{table:model_trained_on_different_supervision}
\end{table}

\noindent\textbf{3) 
Temporal bidirectionality.
}
We compare different FPG loss items in Table~\ref{tab:ablation_localization_loss_func}. 
The baseline is the best model (Row 4) in Table~\ref{table:ablation_study_feature_learning}.
\texttt{S\&E} refers to the localization loss based on start-end boundary probabilities, while \texttt{Content}: further incorporates content-based temporal supervision. \texttt{Flip}: simultaneous forward and backward localization during training. 
The results show that \texttt{S\&E} improves recall, and \texttt{Content} yields extra gains. 
\texttt{Flip} further lifts AP@0.95 by 50\%. 

\begin{table}[t]
\centering
\fontsize{9}{11}\selectfont
\setlength{\tabcolsep}{1mm}
\begin{tabular}{
>{\centering\arraybackslash}m{0.7cm}>{\centering\arraybackslash}m{1cm}
>{\centering\arraybackslash}m{0.7cm}>{\centering\arraybackslash}m{0.7cm}
>{\centering\arraybackslash}m{0.7cm}>{\centering\arraybackslash}m{0.7cm}
>{\centering\arraybackslash}m{0.7cm}>{\centering\arraybackslash}m{0.7cm}>{\centering\arraybackslash}m{0.7cm}}
    \toprule
    \multirow{2}{*}{S \& E} & \multirow{2}{*}{Content} & \multirow{2}{*}{Flip} & \multicolumn{3}{c}{Avg. Precision \%} & \multicolumn{3}{c}{Avg. Recall \%} \\
    \cmidrule(lr){4-6} \cmidrule(lr){7-9}
        &    &    & 0.5 & 0.75 & 0.95 & 50 & 20 & 10 \\
    \midrule
    N & N & N  & 95.81 & 69.82 & 0.66 & 70.41 & 68.44 & 67.61 \\
    
    Y & N  & N  & 95.51 & 86.92 & 6.80 & 87.79 & 87.12 & 86.64 \\
    
    Y & Y & N  & 95.75 & 87.47 & 6.80 & 87.86 & 86.88 & 86.34  \\
    
    Y & Y & Y & \textbf{96.50} & \textbf{92.74} & \textbf{12.71} & \textbf{89.19} & \textbf{88.71} & \textbf{88.29} \\
\bottomrule

\end{tabular}
\caption{Ablation study on the training loss for FPG }
\label{tab:ablation_localization_loss_func}
\end{table}

In Table~\ref{tab:ablation_localization_loss_func}, \texttt{Flip} yields substantial improvements. 
These gains arise from providing complementary boundary cues from bi-directions~\cite{BSN++_Su_Gan_Wu_Qiao_Yan_2021}, enhancing localization under the asymmetry of audio–visual deepfakes. 
Artifacts often manifest differently at the transition of ``real-to-fake'' (onset) and ``fake-to-real'' (offset). Thus, a unidirectional model, which focuses solely on the forward pass, may overlook critical information occurring after these transitions.
By utilizing bidirectional input, the model avoids missing crucial post-transition information and captures additional cues by leveraging both fake-to-real and real-to-fake transitions in the same position, thereby enhancing frame-level discriminability.
This bidirectional perspective not only enriches representations of subtle artifacts but also regularizes the network to produce stable boundaries in both directions.

\begin{table}[t]
    \centering
    \fontsize{9}{11} \selectfont
    \begin{tabular}{>{\centering\arraybackslash}m{0.82cm}
    >{\centering\arraybackslash}m{0.7cm}>{\centering\arraybackslash}m{0.7cm}>{\centering\arraybackslash}m{0.7cm}|>{\centering\arraybackslash}m{0.7cm}>{\centering\arraybackslash}m{0.7cm}>{\centering\arraybackslash}m{0.7cm}}
        \toprule
        \multirow{2}{*}{Scale}  & \multicolumn{3}{c}{Avg. Precision \%} & \multicolumn{3}{c}{Avg. Recall \%} \\
        \cmidrule(lr){2-4} \cmidrule(lr){5-7}
          & 0.5 & 0.75 & 0.95 & 50  & 20 & 10 \\
        \midrule
        0.33x &  88.07 &  83.68 &  10.73 &  88.29 &  87.05 &  85.73 \\
        
        0.50x &  92.63 &  88.75 &  11.13 &  88.75 &  88.06 &  87.43 \\
        
        1x &  96.50  &  92.74 &  12.71 &  89.19 &  88.71 &  88.29 \\
        
        2x &  97.58 &  \textbf{94.57} &  14.31 &  89.04 & 
         88.76 &  88.50  \\
        
        3x &  \textbf{97.88} &  94.52 &  \textbf{14.58} &  \textbf{89.22} &  \textbf{89.01} &  \textbf{88.83} \\ 
        \bottomrule
    \end{tabular}
    \caption{Comparison under different training data set sizes}
    \label{table:model_trained_on_different_data_size}
\end{table}

\begin{figure}[t]
\centering
\includegraphics[width=8cm]{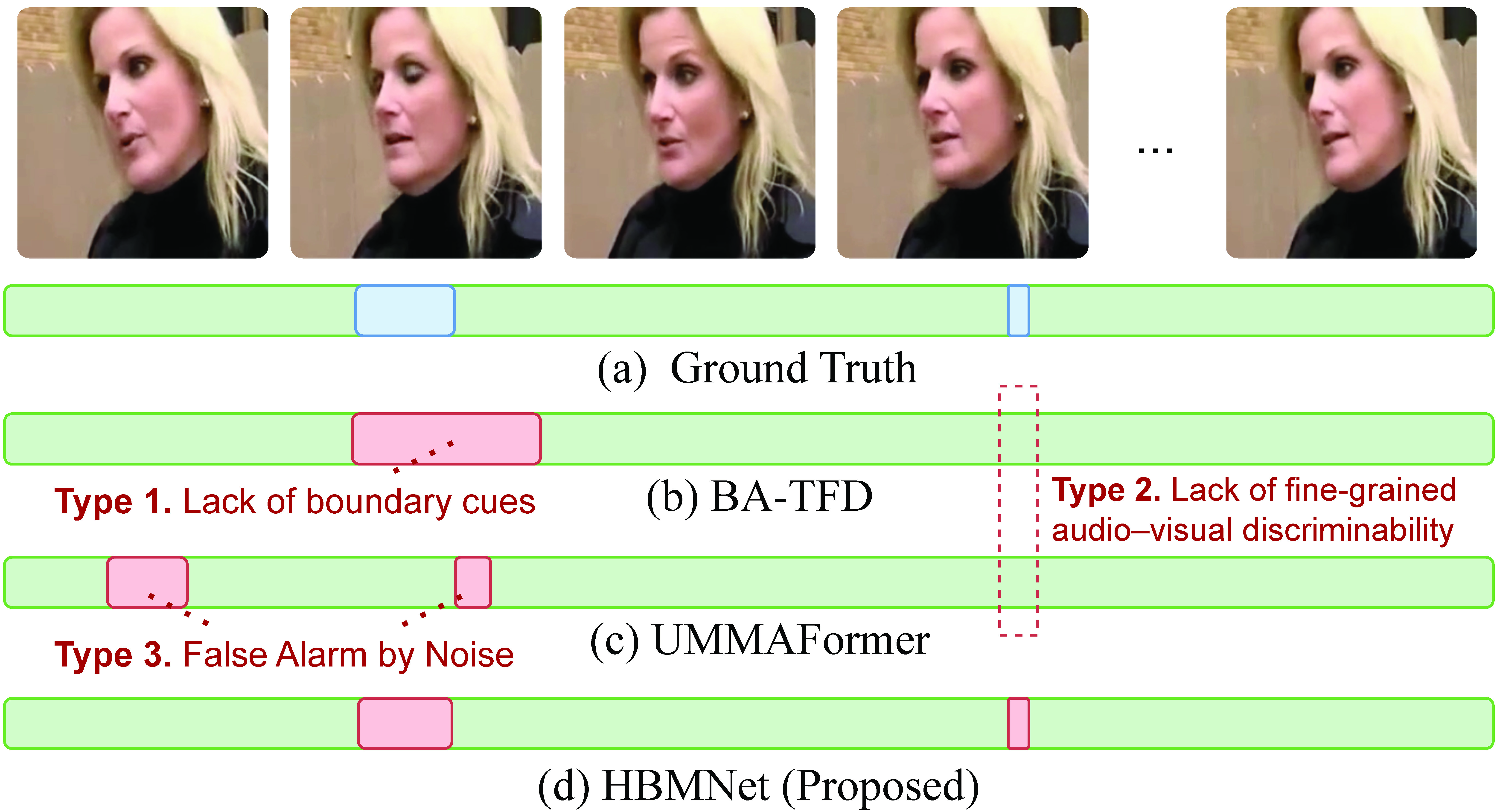}
\caption{Case study on AV-Deepfake-1M. Different colors denote different regions, green and blue denote real and deepfake regions, while red denotes the predicted regions.}
\label{fig:visulizaztion}
\end{figure}

\subsection{Can HBMNet benefit from more data?}
Table~\ref{table:model_trained_on_different_data_size} shows that HBMNet was trained on datasets of varying sizes. 
As the scale of training data increases from \texttt{0.33x} to \texttt{3x} data size, the model improves consistently in precision and recall. 
Precision steadily improves at all IoU thresholds, from 88.07 AP@0.5 and 10.73 AP@0.95 at the \texttt{0.33x} data scale to 97.88 and 14.58 at the \texttt{3x} data scale. Similarly, recall improves across all thresholds, with the largest gains observed between \texttt{0.33x} and \texttt{1x} data scales. 
These trends suggest that HBMNet benefits from additional training data and exhibits potential scalability. 

\subsection{Can HBMNet capture fragmented deepfakes?}

Figure~\ref{fig:visulizaztion} illustrates and compares the models’ predictions, identifying three errors that may cause the performance gap. 
First, insufficient boundary cues \cite{grad-cam_liu24m_interspeech}: although the model can roughly localize deepfake segments, it cannot pinpoint exact boundaries. 
Second, limited fine-grained audio–visual discriminability \cite{cai2024av}: the model often fails to detect very short manipulated fragments. Third, overemphasis on boundary signals at the expense of content context \cite{su2023multi}: by neglecting the actual audio-visual content, the model is prone to spurious false positives caused by background noise.

We observed that BA-TFD exhibits both type 1 and type 2 errors, while UMMAFormer is afflicted by type 1 and type 3 errors; by contrast, HBMNet accurately localizes even the most fragmented deepfake segments, benefiting from three components: frame-level audio–visual supervision, which sharpens discriminability and prevents short manipulated fragments from being overlooked; hierarchical boundary modeling, which integrates coarse proposal-level cues with fine-grained frame-level probabilities to enhance localization precision; and content-aware boundary cues, which use the intrinsic deepfake content to disambiguate, thereby reducing false alarms caused by noise. 

\section{Conclusion}

In this paper, we propose HBMNet, an audio–visual temporal deepfake localization framework that combines an Audio–Visual Feature Encoder, a Coarse Proposal Generator, and a Fine-grained Probabilities Generator to tackle two challenges: insufficient cross-modal encoding and fusion, and limited modeling of hierarchical boundary cues. 
HBMNet enhances audio–visual learning via dedicated encoding and fusion, supported by frame-level supervision to improve discriminability. It leverages multi-scale (proposal- and frame-level) information with bidirectional transitions. Experiments show that encoding and fusion improve precision, while frame-level supervision boosts recall. 
Each module—fusion, temporal scales, and bi-directionality, offers complementary benefits, collectively improving localization. HBMNet outperforms BA-TFD and UMMAFormer and continues to improve with more training data. 

\subsubsection{Acknowledgments.}
This work was supported by the National Science and Technology Council, Taiwan (Grant no. NSTC 112-2634-F-002-005, Advanced Technologies for Designing Trustable AI Services). We also thank the National Center for High-performance Computing (NCHC) of National Applied Research Laboratories (NARLabs) in Taiwan for providing computational and storage resources. 

\bibliography{aaai2026}

\begin{thebibliography}{44}
\providecommand{\natexlab}[1]{#1}

\bibitem[{Afchar et~al.(2018)Afchar, Nozick, Yamagishi, and Echizen}]{Afchar2018meso4}
Afchar, D.; Nozick, V.; Yamagishi, J.; and Echizen, I. 2018.
\newblock MesoNet: a Compact Facial Video Forgery Detection Network.
\newblock In \emph{2018 IEEE International Workshop on Information Forensics and Security (WIFS)}.

\bibitem[{Bodla et~al.(2017)Bodla, Singh, Chellappa, and Davis}]{Bodla2017soft-nms}
Bodla, N.; Singh, B.; Chellappa, R.; and Davis, L.~S. 2017.
\newblock Soft-NMS — Improving Object Detection with One Line of Code.
\newblock In \emph{2017 IEEE International Conference on Computer Vision (ICCV)}.

\bibitem[{Cai et~al.(2024)Cai, Ghosh, Adatia, Hayat, Dhall, Gedeon, and Stefanov}]{cai2024av}
Cai, Z.; Ghosh, S.; Adatia, A.~P.; Hayat, M.; Dhall, A.; Gedeon, T.; and Stefanov, K. 2024.
\newblock AV-Deepfake1M: A Large-Scale LLM-Driven Audio-Visual Deepfake Dataset.
\newblock In \emph{Proceedings of the 32nd ACM International Conference on Multimedia}.

\bibitem[{Cai et~al.(2023)Cai, Ghosh, Dhall, Gedeon, Stefanov, and Hayat}]{ba-tfd+_cai2023glitch}
Cai, Z.; Ghosh, S.; Dhall, A.; Gedeon, T.; Stefanov, K.; and Hayat, M. 2023.
\newblock Glitch in the Matrix: A Large Scale Benchmark for Content Driven Audio-Visual Forgery Detection and Localization.
\newblock \emph{Computer Vision and Image Understanding}, 236.

\bibitem[{Cai et~al.(2022)Cai, Stefanov, Dhall, and Hayat}]{ba-tfd_cai2022you}
Cai, Z.; Stefanov, K.; Dhall, A.; and Hayat, M. 2022.
\newblock Do You Really Mean That? Content Driven Audio-Visual Deepfake Dataset and Multimodal Method for Temporal Forgery Localization.
\newblock In \emph{2022 International Conference on Digital Image Computing: Techniques and Applications (DICTA)}.

\bibitem[{Chen et~al.(2023)Chen, Wu, Meng, Lee, and Jang}]{chen2023push}
Chen, X.; Wu, H.; Meng, H.; Lee, H.-y.; and Jang, J.-S.~R. 2023.
\newblock Push-Pull: Characterizing the Adversarial Robustness for Audio-Visual Active Speaker Detection.
\newblock In \emph{2022 IEEE Spoken Language Technology Workshop (SLT)}.

\bibitem[{Chen et~al.(2024)Chen, Wu, Wang, Lee, and Jang}]{chen2024multi}
Chen, X.; Wu, H.; Wang, C.-C.; Lee, H.-Y.; and Jang, J.-S.~R. 2024.
\newblock Multimodal Transformer Distillation for Audio-Visual Synchronization.
\newblock In \emph{ICASSP 2024 - 2024 IEEE International Conference on Acoustics, Speech and Signal Processing (ICASSP)}.

\bibitem[{Coccomini et~al.(2022)Coccomini, Messina, Gennaro, and Falchi}]{Coccomini2022EfficientViT}
Coccomini, D.~A.; Messina, N.; Gennaro, C.; and Falchi, F. 2022.
\newblock Combining EfficientNet and Vision Transformers for Video Deepfake Detection.
\newblock In \emph{Image Analysis and Processing -- ICIAP 2022}.

\bibitem[{Dosovitskiy et~al.(2021)Dosovitskiy, Beyer, Kolesnikov, Weissenborn, Zhai, Unterthiner, Dehghani, Minderer, Heigold, Gelly, Uszkoreit, and Houlsby}]{dosovitskiy2021an}
Dosovitskiy, A.; Beyer, L.; Kolesnikov, A.; Weissenborn, D.; Zhai, X.; Unterthiner, T.; Dehghani, M.; Minderer, M.; Heigold, G.; Gelly, S.; Uszkoreit, J.; and Houlsby, N. 2021.
\newblock An Image is Worth 16x16 Words: Transformers for Image Recognition at Scale.
\newblock In \emph{International Conference on Learning Representations}.

\bibitem[{Hadsell, Chopra, and LeCun(2006)}]{Hadsell2006contrastive}
Hadsell, R.; Chopra, S.; and LeCun, Y. 2006.
\newblock Dimensionality Reduction by Learning an Invariant Mapping.
\newblock In \emph{2006 IEEE Computer Society Conference on Computer Vision and Pattern Recognition}.

\bibitem[{He et~al.(2025)He, Yi, Tao, Zeng, and Gu}]{he2025manipulatedregionslocalizationpartially}
He, J.; Yi, J.; Tao, J.; Zeng, S.; and Gu, H. 2025.
\newblock Manipulated Regions Localization For Partially Deepfake Audio: A Survey.
\newblock arXiv:2506.14396.

\bibitem[{He et~al.(2016)He, Zhang, Ren, and Sun}]{he2016resnet}
He, K.; Zhang, X.; Ren, S.; and Sun, J. 2016.
\newblock Deep Residual Learning for Image Recognition.
\newblock In \emph{2016 IEEE Conference on Computer Vision and Pattern Recognition (CVPR)}.

\bibitem[{Hu, Shen, and Sun(2018)}]{hu2018senet}
Hu, J.; Shen, L.; and Sun, G. 2018.
\newblock Squeeze-and-Excitation Networks.
\newblock In \emph{2018 IEEE/CVF Conference on Computer Vision and Pattern Recognition}.

\bibitem[{Kinnunen et~al.(2017)Kinnunen, Sahidullah, Delgado, Todisco, Evans et~al.}]{kinnunen17_interspeech}
Kinnunen, T.; Sahidullah, M.; Delgado, H.; Todisco, M.; Evans, N.; et~al. 2017.
\newblock The {ASVspoof} 2017 Challenge: Assessing the Limits of Replay Spoofing Attack Detection.
\newblock In \emph{Proc. INTERSPEECH}.

\bibitem[{Li et~al.(2022)Li, Wu, Fan, Mangalam, Xiong, Malik, and Feichtenhofer}]{li2022mvitv2}
Li, Y.; Wu, C.-Y.; Fan, H.; Mangalam, K.; Xiong, B.; Malik, J.; and Feichtenhofer, C. 2022.
\newblock Mvitv2: Improved multiscale vision transformers for classification and detection.
\newblock In \emph{Proceedings of the IEEE/CVF conference on computer vision and pattern recognition}.

\bibitem[{Lin et~al.(2019)Lin, Liu, Li, Ding, and Wen}]{BMN_Lin_2019_ICCV}
Lin, T.; Liu, X.; Li, X.; Ding, E.; and Wen, S. 2019.
\newblock BMN: Boundary-Matching Network for Temporal Action Proposal Generation.
\newblock In \emph{Proceedings of the IEEE/CVF International Conference on Computer Vision (ICCV)}.

\bibitem[{Lin et~al.(2018)Lin, Zhao, Su, Wang, and Yang}]{BSN_Lin_2018_ECCV}
Lin, T.; Zhao, X.; Su, H.; Wang, C.; and Yang, M. 2018.
\newblock BSN: Boundary Sensitive Network for Temporal Action Proposal Generation.
\newblock In \emph{Proceedings of the European Conference on Computer Vision (ECCV)}.

\bibitem[{Lin et~al.(2017)Lin, Goyal, Girshick, He, and Doll{\'a}r}]{lin2017focal}
Lin, T.-Y.; Goyal, P.; Girshick, R.; He, K.; and Doll{\'a}r, P. 2017.
\newblock Focal loss for dense object detection.
\newblock In \emph{Proceedings of the IEEE international conference on computer vision}, 2980--2988.

\bibitem[{Liu et~al.(2023{\natexlab{a}})Liu, Wang, Qian, and Li}]{liu2023audio}
Liu, M.; Wang, J.; Qian, X.; and Li, H. 2023{\natexlab{a}}.
\newblock Audio-visual temporal forgery detection using embedding-level fusion and multi-dimensional contrastive loss.
\newblock \emph{IEEE Transactions on Circuits and Systems for Video Technology}, 34.

\bibitem[{Liu et~al.(2024)Liu, Zhang, Das, Ma, Tao, and Li}]{grad-cam_liu24m_interspeech}
Liu, T.; Zhang, L.; Das, R.~K.; Ma, Y.; Tao, R.; and Li, H. 2024.
\newblock How Do Neural Spoofing Countermeasures Detect Partially Spoofed Audio?
\newblock In \emph{Proc. INTERSPEECH}.

\bibitem[{Liu et~al.(2023{\natexlab{b}})Liu, Wang, Sahidullah, Patino, Delgado, Kinnunen et~al.}]{liu2022asvspoof2021}
Liu, X.; Wang, X.; Sahidullah, M.; Patino, J.; Delgado, H.; Kinnunen, T.; et~al. 2023{\natexlab{b}}.
\newblock {ASVspoof 2021}: Towards Spoofed and Deepfake Speech Detection in the Wild.
\newblock \emph{IEEE/ACM Trans. Audio, Speech, Lang. Process.}, 31.

\bibitem[{Martinez et~al.(2020)Martinez, Ma, Petridis, and Pantic}]{martinez2020lipreading}
Martinez, B.; Ma, P.; Petridis, S.; and Pantic, M. 2020.
\newblock Lipreading using temporal convolutional networks.
\newblock In \emph{ICASSP 2020-2020 IEEE International Conference on Acoustics, Speech and Signal Processing (ICASSP)}.

\bibitem[{Nautsch et~al.(2021)Nautsch, Wang, Evans, Kinnunen, Vestman, Todisco et~al.}]{nautsch2021asvspoof}
Nautsch, A.; Wang, X.; Evans, N.; Kinnunen, T.~H.; Vestman, V.; Todisco, M.; et~al. 2021.
\newblock {ASVspoof 2019}: spoofing countermeasures for the detection of synthesized, converted and replayed speech.
\newblock \emph{IEEE Trans. Biometrics, Behav., Identity. Sci.}, 3.

\bibitem[{Niizumi et~al.(2021)Niizumi, Takeuchi, Ohishi, Harada, and Kashino}]{niizumi2021byol}
Niizumi, D.; Takeuchi, D.; Ohishi, Y.; Harada, N.; and Kashino, K. 2021.
\newblock Byol for audio: Self-supervised learning for general-purpose audio representation.
\newblock In \emph{2021 International Joint Conference on Neural Networks (IJCNN)}.

\bibitem[{OpenAI et~al.(2024)OpenAI, Achiam, Adler, Agarwal, Ahmad, Akkaya, Aleman, Almeida, Altenschmidt et~al.}]{openai2024gpt4technicalreport}
OpenAI; Achiam, J.; Adler, S.; Agarwal, S.; Ahmad, L.; Akkaya, I.; Aleman, F.~L.; Almeida, D.; Altenschmidt, J.; et~al. 2024.
\newblock GPT-4 Technical Report.
\newblock arXiv:2303.08774.

\bibitem[{P\'{e}rez-Vieites et~al.(2024)P\'{e}rez-Vieites, Moreira-P\'{e}rez, Arag\'{o}n-Kifute, Rom\'{a}n-Sarmiento, and Castro-Gonz\'{a}lez}]{1M-deepfakes-vigo-first}
P\'{e}rez-Vieites, D.; Moreira-P\'{e}rez, J.~J.; Arag\'{o}n-Kifute, A.; Rom\'{a}n-Sarmiento, R.; and Castro-Gonz\'{a}lez, R. 2024.
\newblock Vigo: Audiovisual Fake Detection and Segment Localization.
\newblock In \emph{Proceedings of the 32nd ACM International Conference on Multimedia}.

\bibitem[{Plaquet and Bredin(2023)}]{plaquet23_interspeech}
Plaquet, A.; and Bredin, H. 2023.
\newblock Powerset multi-class cross entropy loss for neural speaker diarization.
\newblock In \emph{Proc. INTERSPEECH}.

\bibitem[{Shi et~al.(2023)Shi, Zhong, Cao, Ma, Li, and Tao}]{Shi_2023_CVPR}
Shi, D.; Zhong, Y.; Cao, Q.; Ma, L.; Li, J.; and Tao, D. 2023.
\newblock TriDet: Temporal Action Detection With Relative Boundary Modeling.
\newblock In \emph{Proceedings of the IEEE/CVF Conference on Computer Vision and Pattern Recognition (CVPR)}.

\bibitem[{Su et~al.(2021)Su, Gan, Wu, Qiao, and Yan}]{BSN++_Su_Gan_Wu_Qiao_Yan_2021}
Su, H.; Gan, W.; Wu, W.; Qiao, Y.; and Yan, J. 2021.
\newblock BSN++: Complementary Boundary Regressor with Scale-Balanced Relation Modeling for Temporal Action Proposal Generation.
\newblock \emph{Proceedings of the AAAI Conference on Artificial Intelligence}, 35(3).

\bibitem[{Su, Wang, and Wang(2023)}]{su2023multi}
Su, T.; Wang, H.; and Wang, L. 2023.
\newblock Multi-level content-aware boundary detection for temporal action proposal generation.
\newblock \emph{IEEE Transactions on Image Processing}, 32.

\bibitem[{Todisco et~al.(2019)Todisco, Wang, Vestman, Sahidullah, Delgado, Nautsch et~al.}]{todisco2019asvspoof}
Todisco, M.; Wang, X.; Vestman, V.; Sahidullah, M.; Delgado, H.; Nautsch, A.; et~al. 2019.
\newblock {ASVspoof 2019}: Future Horizons in Spoofed and Fake Audio Detection.
\newblock In \emph{Proc. INTERSPEECH}.

\bibitem[{Vaswani et~al.(2017)Vaswani, Shazeer, Parmar, Uszkoreit, Jones, Gomez, Kaiser, and Polosukhin}]{vaswani2017attention}
Vaswani, A.; Shazeer, N.; Parmar, N.; Uszkoreit, J.; Jones, L.; Gomez, A.~N.; Kaiser, {\L}.; and Polosukhin, I. 2017.
\newblock Attention is all you need.
\newblock \emph{Advances in neural information processing systems}, 30.

\bibitem[{Vila\c{c}a, Yu, and Viana(2025)}]{Vila2025avlearning}
Vila\c{c}a, L.; Yu, Y.; and Viana, P. 2025.
\newblock A Survey of Recent Advances and Challenges in Deep Audio-Visual Correlation Learning.
\newblock \emph{ACM Comput. Surv.}, 57(12).

\bibitem[{Wang et~al.(2024{\natexlab{a}})Wang, Zhao, Yang, Long, and Li}]{Wang2024tal}
Wang, B.; Zhao, Y.; Yang, L.; Long, T.; and Li, X. 2024{\natexlab{a}}.
\newblock Temporal Action Localization in the Deep Learning Era: A Survey.
\newblock \emph{IEEE Transactions on Pattern Analysis and Machine Intelligence}, 46(4): 2171--2190.

\bibitem[{Wang et~al.(2016)Wang, Xiong, Wang, Qiao, Lin, Tang, and Van~Gool}]{wang2016temporal}
Wang, L.; Xiong, Y.; Wang, Z.; Qiao, Y.; Lin, D.; Tang, X.; and Van~Gool, L. 2016.
\newblock Temporal segment networks: Towards good practices for deep action recognition.
\newblock In \emph{European conference on computer vision}, 20--36. Springer.

\bibitem[{Wang et~al.(2024{\natexlab{b}})Wang, Delgado, Tak, Jung, Shim, Todisco et~al.}]{Wang2024_ASVspoof5}
Wang, X.; Delgado, H.; Tak, H.; Jung, J.-w.; Shim, H.-j.; Todisco, M.; et~al. 2024{\natexlab{b}}.
\newblock {ASVspoof 5}: {Crowdsourced} Speech Data, Deepfakes, and Adversarial Attacks at Scale.
\newblock In \emph{Proc. ASVspoof Workshop}.

\bibitem[{Wu et~al.(2015)Wu, Kinnunen, Evans, Yamagishi, Hanilçi, Sahidullah, and Sizov}]{wu15e_interspeech}
Wu, Z.; Kinnunen, T.; Evans, N.; Yamagishi, J.; Hanilçi, C.; Sahidullah, M.; and Sizov, A. 2015.
\newblock {ASVspoof 2015}: the first automatic speaker verification spoofing and countermeasures challenge.
\newblock In \emph{Proc. INTERSPEECH}.

\bibitem[{Yi et~al.(2022)Yi, Fu, Tao, Nie, Ma, Wang, Wang et~al.}]{yi2022add}
Yi, J.; Fu, R.; Tao, J.; Nie, S.; Ma, H.; Wang, C.; Wang, T.; et~al. 2022.
\newblock Add 2022: the first audio deep synthesis detection challenge.
\newblock In \emph{Proc. IEEE Int. Conf. Acoust., Speech, Signal Process.}

\bibitem[{Yi et~al.(2023)Yi, Tao, Fu, Yan, Wang, Wang et~al.}]{yi2023add}
Yi, J.; Tao, J.; Fu, R.; Yan, X.; Wang, C.; Wang, T.; et~al. 2023.
\newblock ADD 2023: the Second Audio Deepfake Detection Challenge.
\newblock In \emph{DADA@IJCAI}.

\bibitem[{Zhang, Wu, and Li(2022)}]{Zhang2022Actionformer}
Zhang, C.-L.; Wu, J.; and Li, Y. 2022.
\newblock ActionFormer: Localizing Moments of Actions with Transformers.
\newblock In \emph{Computer Vision -- ECCV 2022}.

\bibitem[{Zhang et~al.(2023{\natexlab{a}})Zhang, Wang, Cooper, Evans, and Yamagishi}]{lin2023partialspoof}
Zhang, L.; Wang, X.; Cooper, E.; Evans, N.; and Yamagishi, J. 2023{\natexlab{a}}.
\newblock The PartialSpoof Database and Countermeasures for the Detection of Short Fake Speech Segments Embedded in an Utterance.
\newblock \emph{IEEE/ACM Transactions on Audio, Speech, and Language Processing}, 31.

\bibitem[{Zhang et~al.(2023{\natexlab{b}})Zhang, Wang, Du, Liu, Zhou, and Zeng}]{zhang2023ummaformer}
Zhang, R.; Wang, H.; Du, M.; Liu, H.; Zhou, Y.; and Zeng, Q. 2023{\natexlab{b}}.
\newblock UMMAFormer: {A} Universal Multimodal-adaptive Transformer Framework for Temporal Forgery Localization.
\newblock In \emph{Proceedings of the 31st {ACM} International Conference on Multimedia}.

\bibitem[{Zhang et~al.(2024)Zhang, Miao, Luo et~al.}]{1M-deepfakes-mfms-second}
Zhang, Y.; Miao, C.; Luo, M.; et~al. 2024.
\newblock MFMS: Learning Modality-Fused and Modality-Specific Features for Deepfake Detection and Localization Tasks.
\newblock In \emph{Proceedings of the 32nd ACM International Conference on Multimedia}.

\bibitem[{Zhou et~al.(2018)Zhou, Rahman~Siddiquee, Tajbakhsh, and Liang}]{zhou2018unet++}
Zhou, Z.; Rahman~Siddiquee, M.~M.; Tajbakhsh, N.; and Liang, J. 2018.
\newblock Unet++: A nested u-net architecture for medical image segmentation.
\newblock In \emph{International workshop on deep learning in medical image analysis}.

\end{thebibliography}

\end{document}